\documentstyle[11pt,epsf]{article}

\catcode`\@=11
\long\def\@makefntext#1{
\protect\noindent \hbox to 3.2pt {\hskip-.9pt
$^{{\ninerm\@thefnmark}}$\hfil}#1\hfill}		

\def\@makefnmark{\hbox to 0pt{$^{\@thefnmark}$\hss}}  

\def\ps@myheadings{\let\@mkboth\@gobbletwo
\def\@oddhead{\hbox{}
\rightmark\hfil\ninerm\thepage}
\def\@oddfoot{}\def\@evenhead{\ninerm\thepage\hfil
\leftmark\hbox{}}\def\@evenfoot{}
\def\sectionmark##1{}\def\subsectionmark##1{}}

\renewcommand{\thefootnote}{\fnsymbol{footnote}}

\newcounter{sectionc}\newcounter{subsectionc}\newcounter{subsubsectionc}
\renewcommand{\section}[1] {\vspace*{0.8cm}\addtocounter{sectionc}{1}
\setcounter{subsectionc}{0}\setcounter{subsubsectionc}{0}\noindent
        {\normalsize\bf\thesectionc. #1}\par\vspace*{0.4cm}}
\renewcommand{\subsection}[1] {\vspace*{0.6cm}\addtocounter{subsectionc}{1}
	\setcounter{subsubsectionc}{0}\noindent
	{\normalsize\it\thesectionc.\thesubsectionc. #1}\par\vspace*{0.4cm}}
\renewcommand{\subsubsection}[1]
{\vspace*{0.6cm}\addtocounter{subsubsectionc}{1}
	\noindent {\normalsize\rm\thesectionc.\thesubsectionc.\thesubsubsectionc.
	#1}\par\vspace*{0.4cm}}

\def\abstracts#1{{

\centering{\begin{minipage}{12.2truecm}\footnotesize\baselineskip=12pt\noindent
	\centerline{\footnotesize ABSTRACT}\vspace*{0.3cm}
	\parindent=0pt #1
	\end{minipage}}\par}}

\renewenvironment{thebibliography}[1]
	{\begin{list}{\arabic{enumi}.}
	{\usecounter{enumi}\setlength{\parsep}{0pt}
\setlength{\leftmargin 1.25cm}{\rightmargin 0pt}
	 \setlength{\itemsep}{0pt} \settowidth
	{\labelwidth}{#1.}\sloppy}}{\end{list}}

\newcounter{itemlistc}
\newcounter{romanlistc}
\newcounter{alphlistc}
\newcounter{arabiclistc}

 1
 1
 1

\font\ninerm=cmr9

\newcommand{\cZ}{\cal{Z}}
\newcommand{\cU}{{\cal U}}
\newcommand{\cD}{{\cal D}}
\newcommand{\Z}{{Z \!\!\! Z}}
\newcommand{\eq}[1]{(\ref{#1})}
\newcommand{\dual}{\mbox{}^{\ast}}

\newcommand{\CK}[1]{\mbox{\scriptsize c}_{\mbox{$\scriptstyle #1$}}}
\newcommand{\beq}{\begin{equation}}
\newcommand{\eeq}{\end{equation}}
\newcommand{\beqn}{\begin{eqnarray}}
\newcommand{\eeqn}{\end{eqnarray}}
\newcommand{\nsum}[2]{\sum_{ #1(\CK{#2}) \in \Z }}

\newcommand{\intpi}{\int\limits_{-\pi}^{+\pi} {\cD}}
\newcommand{\intinf}{\int\limits_{-\infty}^{+\infty} {\cD}}
\newcommand{\dd}{\mbox{d}}
\newcommand{\expb}[1]{\exp\left\{ #1 \right\} }

\date{}
\begin{document}
\rightline{\normalsize ITEP--TH--14/95}
\rightline{\normalsize hep--lat/9512030}
\vspace{1cm}
\centerline{\normalsize\bf MONOPOLE ORDER PARAMETER}
\baselineskip=22pt
\centerline{\normalsize\bf IN $SU(2)$ LATTICE GAUGE THEORY\footnote{
Talk given at 29--th International Symposium on the Theory
of Elementary Particles, Buckov, 29 August - 2 September, 1995.
}
}
\vspace*{0.6cm}
\centerline{\footnotesize M.N.~CHERNODUB, M.I.~POLIKARPOV and A.I.~VESELOV}
\baselineskip=13pt
\vspace*{0.3cm}
\centerline{\footnotesize\it ITEP, B.Cheremushkinskaya 25, Moscow,
117259, Russia}
\vspace*{0.8cm}

\abstracts{
We present the results of the numerical calculation of the probability
distribution of the value of the monopole creation operator in $SU(2)$
lattice gluodynamics. We work in the maximal abelian projection. It
occurs that at the low temperature, below the deconfinement phase
transition the maximum of the distribution is shifted from zero, which
means that the effective constraint potential is of the Higgs
type. Above the phase transition the minimum of the potential (the
maximum of the monopole field distribution) is at the zero value of
the monopole field. This is the direct proof of the existence of the
abelian monopole condensate in the confinement phase of the
gluodynamics, which confirms the dual superconductor model of the
confining vacuum.
}

\vspace*{0.6cm}
\normalsize\baselineskip=15pt
\setcounter{footnote}{0}
\renewcommand{\thefootnote}{\alph{footnote}}

\section{Introduction}

The monopole mechanism of the colour confinement is generally accepted
by the community. Still there are many open questions. In the lattice
gluodynamics it is very important to find the order parameter,
constructed from the monopole field, for the deconfinement phase
transition. The first candidate is the monopole condensate, which
should be nonzero in the confinement phase and vanish at the phase
transition. To study the monopole condensate we need the explicit
expression for the operator $\Phi_{mon}(x)$, which creates the abelian
monopole at the point $x$. The operator $\Phi_{mon}(x)$ was found for
the compact electrodynamics with the Villain form of the action by
Fr\"ohlich and Marchetti \cite{FrMa87}, and it was studied numerically
in refs.\cite{Wiese}.  In Section 2 we construct the monopole creation
operator for an arbitrary abelian projection of lattice $SU(2)$
gluodynamics. The numerical results presented in Section 3 are
obtained for the maximal abelian projection, for this projection many
numerical simulations show that the gluodynamic vacuum behaves as the
dual superconductor (see \cite{Suz93} and references therein). In
refs.\cite{DiGi} the another form of the monopole creation operator
was studied, and it was found that its expectation value vanishes in
the deconfinement phase; as we discuss at the end of Section 2, our
operator is positively defined, therefore our definition of
$\Phi_{mon}(x)$ differs from that of ref.\cite{DiGi}, still our
conclusions and that of ref.\cite{DiGi} coincide: the monopole
condensate exists in the confinement phase of lattice gluodynamics.
The analogous claim is done in ref.\cite{IvPoPo93}, where the monopole
condensate is calculated on the basis of the percolation properties of
the monopole currents.

\section{Monopole Creation Operator}

First we give the formal construction of the monopole creation operator
for the abelian projection of $SU(2)$ gluodynamics.
Let us parametrize $SU(2)$ link matrix in the standard way:  $U^{11}_{x\mu}
= \cos \phi_{x\mu}\, e^{i\theta_{x\mu}}; \ U^{12}_{x\mu} = \sin
\phi_{x\mu}\, e^{i\chi_{x\mu}};$ $\ U^{22}_{x\mu} = U^{11 *}_{x\mu}; \
U^{21}_{x\mu} = - U^{12 *}_{x\mu};$ $\ 0 \le \phi \le \pi/2, \ -\pi <
\theta,\chi \le \pi$.

The plaquette action in terms of the angles $\phi, \ \theta $ and
$\chi$ can be written as follows:

\beq
S_P  =  \frac{1}{2}\mbox{Tr}\, U_1 U_2 U_3^+ U_4^+ = S^a + S^n + S^i\;,
\label{SP}
\eeq
where

\beq
S^a  = \cos \theta_P\,
\cos\phi_1\, \cos\phi_2\, \cos\phi_3\, \cos\phi_4,
\label{Sa} \\
\eeq
$S^n$ and $S^i$ describe the interaction of the fields $\theta$ and $\chi$
and selfinteraction of the field $\chi$~\cite{ChPoVe95};

\beq
\theta_P  = \theta_1 + \theta_2 - \theta_3 - \theta_4\;, \label{P}
\eeq
here the subscripts $1,...,4$ correspond to the links of the plaquette:  $1
\rightarrow \{x,x+\hat{\mu}\},...,4 \rightarrow \{x,x+\hat{\nu}$\}.

If we fix the abelian projection,
each term $S^a$, $S^n$ and $S^i$ is invariant under the residual
$U(1)$ gauge transformations:

\beqn
    \theta_{x\mu} & \to & \theta_{x\mu} +\alpha_x -\alpha_{x+\hat{\mu}}
     \label{u1th}\;,\\
    \chi_{x\mu} & \to & \chi_{x\mu} +\alpha_x + \alpha_{x+\hat{\mu}}\;.
      \label{u1chi}
\eeqn

The operator which creates the monopoles at the point $x$ of the dual
lattice is defined as follows:

\beqn
   \cU(x) = \exp \left\{ \beta [ - S(\theta_P,\phi) + S(\theta_P +
   W_P(x),\phi)] \right\}\,,  \label{Ux1}
\eeqn
we define the function $W_P(x)$ below.
Substituting eq.(\ref{SP}--\ref{Sa}) in eq.\eq{Ux1} we get

\beqn
   \cU(x) = \exp \left\{ \sum_P \tilde\beta \left[ - \cos(\theta_P) +
   \cos(\theta_P + W_P(x)) \right] \right\}\,, \label{Ux2}
\eeqn
where
$\tilde\beta = \cos\phi_1 \cos\phi_2 \cos\phi_3\cos\phi_4 \,\beta$.
Effectively the monopole creation operator shifts all abelian
plaquette angles $\theta_P$.

For the compact electrodynamics with the Villain type of the action
the above definition coincides with the definition of Fr\"ohlich and
Marchetti \cite{FrMa87}. For the general type of the action in compact
electrodynamics we can use the described above construction. The proof
is given in Appendix A. The gluodynamics in the abelian projection
contains the compact gauge field $\theta$ and the charged {\it
vector} field $\chi$. The action \eq{SP} in terms of the fields $\theta$
and $\chi$ is rather nontrivial, and at the moment we have no
proof that the above construction of the monopole creation operator
is valid in this case. Still for the rather similar Abelian
-- Higgs model, with the general type of the action, the proof exists,
and it is analogous to one given in Appendix A. Moreover the
numerical results, presented in the next section, clearly show that
the suggested operator is the order parameter for the deconfinement
phase transition.

\section{Numerical Results}

We present the results of the numerical calculations on the lattice
$10^3\cdot 4$, we impose the anti--periodic boundary conditions in space
directions, since the construction of the operator $\cU$ can be done
only in the time slice with the anti--periodic boundary conditions.  The
periodic boundary conditions are forbidden due to the Gauss law;
formally there is no solution of equation \eq{divD} in the finite
box with the periodic boundary conditions. To see that we have the
order parameter for the deconfinement phase transition it is very
convenient to study the probability distribution of the operator
$\cU$. It means that we calculate the expectation value $<\delta (\varphi -
\cU(x))>$. The quantity like the effective constraint potential,

\beq
e^{-V_{eff}(\varphi)} = <\delta (\varphi - \frac{1}{V}\sum_x \cU(x))>
\eeq
has more physical meaning than the probability distribution. At the
moment we have no enough statistics to calculate $V_{eff}(\varphi)$,
and we present our results for the quantity $V(\varphi)$, defined as:

\beq
e^{-V(\varphi)} = <\delta (\varphi - \cU(x))>.
\eeq

In Figs. 1(a) and 1(b) we show $V(\varphi)$ for the confinement and the
deconfinement phases. It is clearly seen that in the confinement phase
the minimum of $V(\varphi)$ is shifted from zero, while in the
deconfinement phase the minimum is at zero value of the monopole field
$\phi$. We used the positively definite operator $\cU(x)$ \eq{Ux2}, but
in the dual representation the creation operator of the monopole
\eq{Ud} is not positively defined, the sign is loosed since we perform
the inverse duality transformation on the infinite lattice. On the
finite lattice it is possible to get the non--positive defined operator
$\cU(x)$, in that case instead of Fig. 1(a) we get the Higgs --
type potential. This little bit more complicated calculations are now
in progress. Still Figs.~1(a),(b) clearly show that the position of the
minimum of $V(\varphi)$ plays the role of the order parameter. On Fig.2
we show the dependence of the position of the minimum, $\varphi_c$, on
the temperature, it is seen that $\varphi_c$ vanishes at the point of the
phase transition.

\newpage
\noindent
{\normalsize\bf Acknowledgments}\par\vspace*{0.4cm}

Authors are grateful to P.~van Baal, A.~Di~Giacomo, R.~Haymaker,
T.~Ivanenko, Y.~Matsubara, A.~Pochinsky, T.~Suzuki and U.~-J.~Wiese
for useful discussions. This work was supported by the grant number
MJM300, financed by the International Science Foundation, by the JSPS
Program on Japan -- FSU scientists collaboration, by the Grant
INTAS-94-0840, and by the grant number 93-02-03609, financed by the
Russian Foundation for the Fundamental Sciences.

\vspace*{0.8cm}
\noindent
{\normalsize\bf Appendix A}\par\vspace*{0.4cm}

Below we construct the monopole creation operator for the compact
electrodynamics with the general type of the action; the similar
construction exists for the compact Abelian--Higgs model with the
general type of the action. First we perform the duality
transformation of the partition function for the $4D$ lattice compact
electrodynamics:

\beq
 \cZ = \intpi \theta
         \expb{ -S(\dd\theta)}, \label{ZQED}
\eeq

We use the notations of the calculus of differential forms on the
lattice \cite{BeJo82} (see also Appendix B). The symbol
$\int\cD\theta$ denotes the integral over all link variables $\theta$.
The partition function of the dual theory,

\beqn
{\cZ}^d = \sum_{ {\scriptstyle \dual n(\CK{1}) \in \Z}  }
\expb{- \dual S(\dd \dual n)},\\
\dual S(p)=
- \ln \intpi F\expb{-S(F)+ i(F,p)},
\eeqn
can be represented as the following limit of the partition
function for the Abelian--Higgs theory:

\beqn
{\cZ}^d = \lim_{\kappa \rightarrow\infty}
\intpi \dual \varphi\intinf \dual B
\sum_{ \scriptstyle {\dual n(\CK{1}) \in \Z}}
\exp\bigl\{-\dual S(\dd \dual B/2\pi) -
\kappa \|\dual B - \dd\dual\varphi + 2 \pi \dual n\|^2 \bigr\},
\label{AH2}
\eeqn
here $\dual S(\dd \dual B/2\pi)$ is the kinetic energy of the dual
gauge field $\dual B$ (the analogue of ${\tilde F}_{\mu\nu}^2$) and
the Higgs field $\exp\{i\, \dual\varphi\}$ carry magnetic charge, since
it interacts via the covariant derivative with the dual gauge field
$\dual B$. The Dirac operator \cite{Dir55},

\beqn
{\cU}^d(x) = e^{i\dual\varphi}\cdot
\expb{-i(\dual D_x,\dual B)}, \label{Ud} \\
\delta \dual D_x = \dual \delta_x \label{divD}
\eeqn
is the gauge invariant monopole creation operator. It creates the
cloud of photons and the monopole at the point $x$. In \eq{divD}
$\dual \delta_x$ is the lattice $\delta$--function, it equals to unity
at the cite $x$ of the dual lattice and is zero at other cites. Note
that in the above formulas the radial part of the Higgs field which
carry the magnetic charge is fixed to unity.

Coming back to the original partition function \eq{ZQED} we get the
expectation value of the monopole creation operator in terms of the
fields $\theta$:

\beqn
<\cU(x)> =\frac{1}{\cZ} \intpi\theta
         \expb{ -S(\dd\theta  + W_P(x))},\\
W_P(x) = 2 \pi \delta\Delta^{-1}(D_x-\omega_x)),
\eeqn
where the Dirac string attached to the monopole \cite{FrMa87}, is
represented by the integer valued 1-form $\dual \omega_x$, which
satisfies the equation: $\delta \dual \omega_x = \dual \delta_x$.

The partition function \eq{ZQED} for the electrodynamics with the
general type of the action can be represented as the sum over the
monopole closed currents \cite{ChPoVe952}. It is straightforward to
show that the monopole creation operator $\cU(x)$ creates the
not--closed monopole trajectory which starts at the point $x$. This
fact shows that $\cU(x)$ is the monopole creation operator.

\vspace*{0.8cm}
\noindent
{\normalsize\bf Appendix B}\par\vspace*{0.4cm}

Below we briefly summarize the main notions from the theory of
differential forms on the lattice \cite{BeJo82}.  The advantages of the
calculus of differential forms consists in the general character of the
expressions obtained. Most of the transformations depend neither on the
space--time dimension, nor on the rank of the fields. With minor
modifications, the transformations are valid for lattices of any form
(triangular, hypercubic, random, {\it etc}). A differential form of rank $k$ on
the lattice is a function $\phi_{k}$ defined on $k$-dimensional cells $c_k$
of the lattice, {\it eg} the scalar (gauge) field is a 0--form (1--form). The
exterior differential operator {\it d} is defined as follows:

\beq
(\dd \phi ) (c_{k+1}) =\sum_{\CK{k} \in \partial\CK{k+1}} \phi(c_{k}).
\label{def-dd}
\eeq
Here $\partial c_{k}$ is the oriented boundary of the $k$-cell
$c_{k}$. Thus the operator {\it d} increases the rank of the form by unity;
$\dd \varphi$ is the link
variable constructed, as usual, in terms of the site angles $\varphi$, and
$\dd A$ is the plaquette variable constructed from the link variables $A$.
The scalar product is defined in the standard way:
if $\varphi$ and $\psi$ are $k$-forms, then
$(\varphi,\psi)=\sum_{c_k}\varphi(c_k)\psi(c_k)$, where $\sum_{c_k}$ is the
sum over all cells $c_k$.
To any $k$--form on the $D$--dimensional lattice there
corresponds a $(D-k)$--form $\dual\Phi(\dual c_k)$ on the dual lattice,
$\dual c_k$ being the $(D-k)$--dimensional cell on the dual lattice. The
codifferential $\delta=\dual \dd \dual$ satisfies the partial
integration rule: $(\varphi,\delta\psi)=(\dd\varphi,\psi)$.
Note that $\delta \Phi(c_k)$ is a $(k-1)$--form and
$\delta \Phi(c_0) = 0$. The norm is defined by: $\|a\|^2=(a,a)$; therefore,
$\|B - \dd\varphi+2\pi n\|^2$ in \eq{AH2} implies summation over all links.
$\nsum{l}{1}$ denotes the sum over all configurations of the integers $l$
attached to the links $c_1$. The action \eq{AH2} is invariant under the
gauge transformations $B' = B + \dd \alpha$, $\varphi' = \varphi + \alpha$
due to the well known property $\dd^2 = \delta^2 = 0$. The
lattice Laplacian is defined by: $\Delta = \dd\delta + \delta\dd$.

\vspace*{0.8cm}
\noindent
{\normalsize\bf Figure Captions}\par\vspace*{0.4cm}

Fig1. $V(\varphi)$ for the confinement~(a) and the deconfinement~(b) phases.

Fig2. Position of the minimum $\varphi_c$ of $V(\varphi)$ {\it vs.}
temperature $T$.

\vspace*{1.0cm}
\noindent
{\normalsize\bf References}\par

\begin{figure}[ht]
   \epsfxsize=18cm
   \centerline{\epsffile{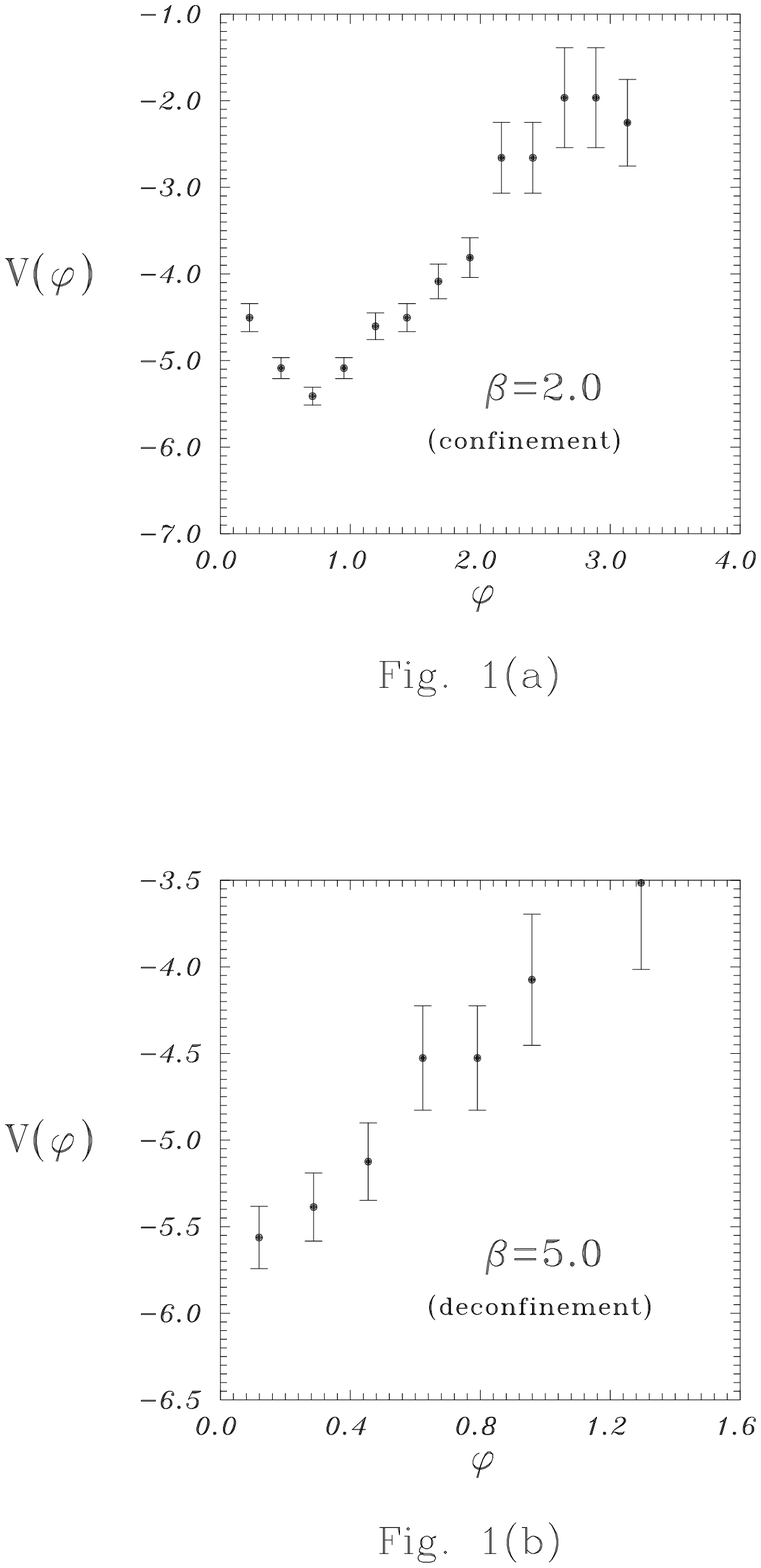}}
\end{figure}

~
\vspace{1cm}

\begin{figure}[ht]
   \epsfxsize=18cm
   \centerline{\epsffile{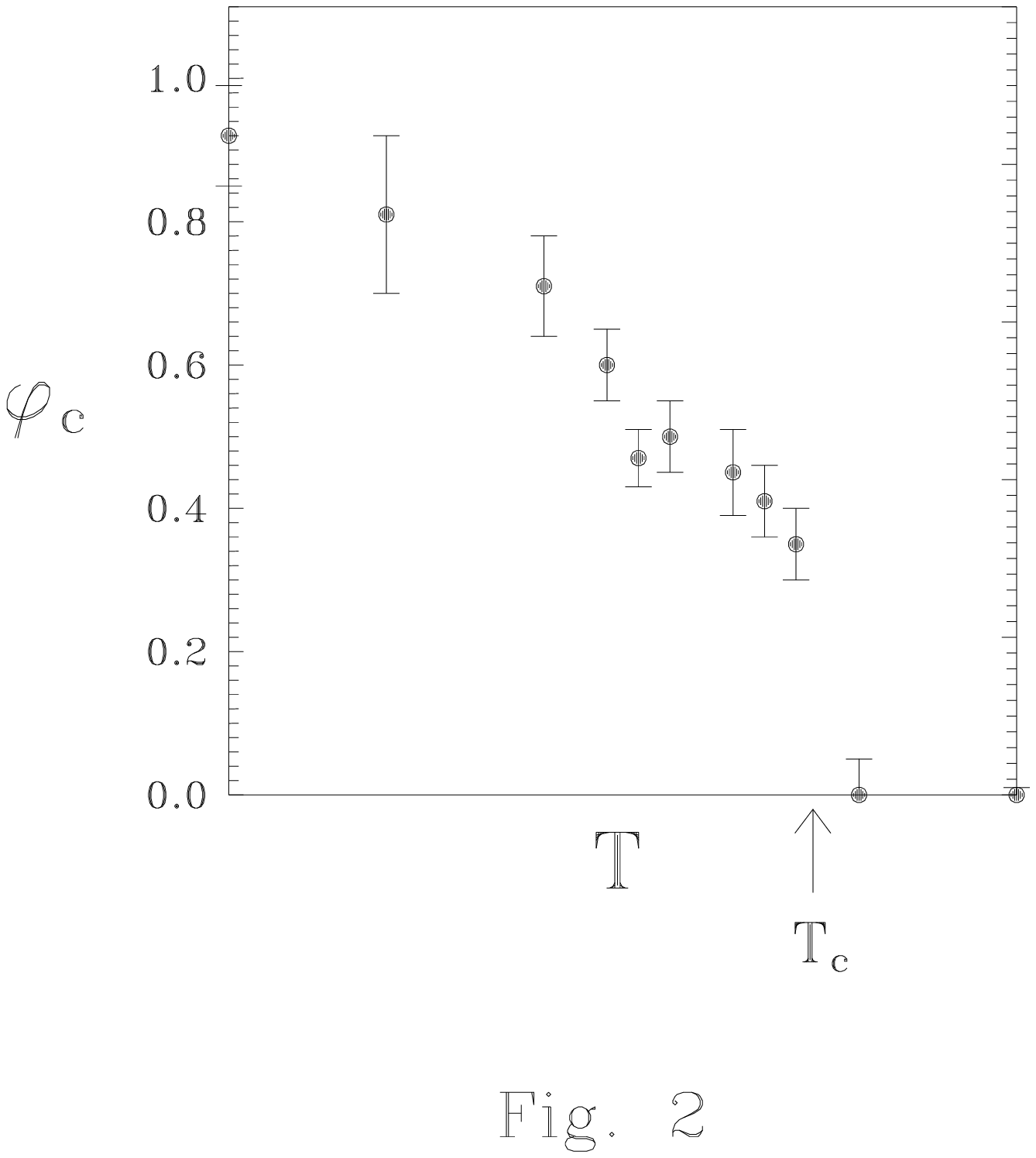}}
\end{figure}

\end{document}